\documentclass{webofc}
\usepackage[varg]{txfonts}   
\usepackage{amsmath}
\usepackage{booktabs}
\usepackage{capt-of}
\usepackage{hyperref}
\usepackage{url}
\hypersetup{colorlinks=true,citecolor=blue,urlcolor=blue,linkcolor=blue}

\newcommand{\alem}{\alpha_{\mathrm{em}}}
\newcommand{\amu}{a_\mu}
\newcommand{\tenten}{\times10^{-10}}
\newcommand{\dd}{\mathrm{d}}
\DeclareMathOperator{\im}{Im}

\begin{document}

\title{Electromagnetic isospin breaking in hadronic vacuum\\ polarization
within a VMD model}

\author{\firstname{Volodymyr} \lastname{Biloshytskyi}\inst{1}\fnsep\thanks{\email{volodymyr.biloshytskyi@univ-amu.fr}}}

\institute{Aix-Marseille Universit\'e, Universit\'e de Toulon, CNRS, CPT, Marseille, France}

\abstract{
One of the challenging components of modern lattice-QCD computations of the leading hadronic vacuum polarization contribution to the muon $(g-2)_\mu$ is the isospin-breaking correction due to electromagnetism. Dominated by long distances, this correction suffers from severe signal-to-noise issues and finite-volume effects. To provide infinite-volume benchmarks for various quark-level Wick contractions, we estimate their sizes phenomenologically at the physical point, using a previously developed vector-meson-dominance model. The pion contributions in this model sum to around $-7.1\times 10^{-10}$, dominated by the electromagnetic pion mass shift. The ultraviolet-finite $(2+2)$a (``dumbbell'') contraction accounts for approximately two thirds of this model sum; for both, about 30\% comes from Euclidean times beyond 2.8~fm.
}

\maketitle

\section{Introduction}
\label{sec:intro}

The muon anomalous magnetic moment $\amu\equiv (g-2)_\mu/2$ is measured with a precision of 124~ppb,
$\Delta\amu^{\rm exp}\simeq1.5\tenten$~\cite{Muong-2:2025xyk,Muong-2:2006rrc}. The leading hadronic vacuum
polarization (HVP) contribution dominates the uncertainty of the Standard-Model prediction in
the 2025 Theory Initiative update~\cite{Aliberti:2025beg}.\footnote{HVP also enters precision atomic
spectroscopy~\cite{Hagelstein:2026mek} and lattice determinations of hadronic contributions to the running of the electroweak gauge couplings~\cite{Conigli:2025qvh,Conigli:2026stl}.}
In view of tensions among data-driven evaluations based on $e^+e^-$
data~\cite{CMD-3:2023alj,Hoferichter:2026juw}, that update takes the HVP from lattice
QCD~\cite{Borsanyi:2020mff,Ce:2022kxy,Djukanovic:2024cmq,RBC:2024fic,MILC:2024ryz,FermilabLatticeHPQCD:2024ppc};
for recent BMW/DMZ results, see Refs.~\cite{Boccaletti:2024guq,BMW:2026qvl}. Hadronic $\tau$ decays offer an alternative
input, once corrected for isospin breaking~\cite{Allen:2026iad,Bruno:2026yba,Hoferichter:2026tqi}.

Lattice calculations commonly incorporate isospin-breaking effects through a perturbative expansion
about the isospin-symmetric point~\cite{deDivitiis:2013xla,Giusti:2019xct,RC:2025zoa},
\begin{equation}
\amu^{\rm HVP}=\amu^{\rm HVP}\big|_{\rm iso}
+\delta m\,\frac{\partial\amu^{\rm HVP}}{\partial\,\delta m}\bigg|_{\rm iso}
+\alem\,\frac{\partial\amu^{\rm HVP}}{\partial\alem}\bigg|_{\rm iso}+\dots,
\qquad \delta m=m_d-m_u ,
\label{eq:IB}
\end{equation}
where $\alem$ is the fine-structure constant, $m_d$ and $m_u$ are the down- and up-quark masses.
The electromagnetic (QED) correction to the HVP, a few units of $10^{-10}$~\cite{RBC:2018dos,Borsanyi:2020mff,Djukanovic:2024cmq,Hoferichter:2023sli},
must be determined with an accuracy of a few tens of percent. Achieving this precision is challenging
because most of the correction comes from long distances, where the signal
deteriorates. This requires reconstructing the tail~\cite{RBC:2018dos,Djukanovic:2024cmq,Parrino:2025afq,Lehner:2025qrl}
or using $e^+e^-$ data for the physical HVP tail, including isospin
breaking~\cite{RBC:2018dos,Boccaletti:2024guq}.
The correction also involves a number of Wick contractions, which are costly to evaluate~\cite{Erb:2026dtl}
and noisy when quark-disconnected~\cite{Parrino:2026zqw}. Furthermore, only some of these contractions are ultraviolet (UV)
finite~\cite{Parrino:2025afq}.

Here we estimate the QED correction to HVP using a vector-meson-dominance (VMD)
model~\cite{Biloshytskyi:2025fjm}.\footnote{Two-pion states, including the $\rho$ resonance,
dominate the HVP beyond about 1~fm in Euclidean time. Chiral perturbation theory describes
the low-energy region, with HVP calculations now reaching three loops~\cite{Lellouch:2025rnz,Lellouch:2026xdr}.}
To connect this hadronic description to lattice Wick
contractions, we use the Cottingham-like formula~\cite{Biloshytskyi:2022ets}, which joins two
currents of the forward light-by-light four-point function with a photon propagator.
This relation underlies lattice studies of the UV-finite dumbbell~\cite{Parrino:2025afq}
and of isospin-violating HVP at the SU(3)-symmetric point~\cite{Erb:2025nxk}.
Leading-order (LO) partially quenched chiral perturbation theory (PQChPT) then supplies the
charge factors for individual contractions~\cite{Bijnens:2016hgx,Chao:2021tvp,Parrino:2025afq}.

The model uses an effective VMD Lagrangian of pions, $\rho$ and $\omega$ mesons, photons and
nucleons~\cite{Kroll:1967it,Gounaris:1968mw,Klingl:1996by}, with parameters fixed by experiment.
In the heavy-nucleon limit, nucleon loops generate the anomalous vertices and fix the isospin
structure of the form factors.

\begin{figure}[ht]
\centering
\includegraphics[width=0.50\textwidth,clip]{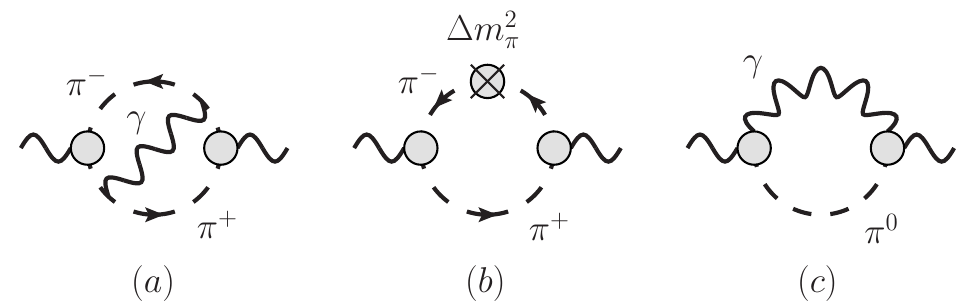}
\caption{The three pion contributions retained in the model~\cite{Biloshytskyi:2025fjm}:
(a) photon correction to the $\pi^+\pi^-$ loop, (b) conversion term due to the charged-pion mass shift $\Delta m_\pi^2$,
(c) $\pi^0\gamma$ loop. Shaded blobs: couplings to the external photon, dressed by the $\rho$ and, for (c), also by
$\omega$ mesons.}
\label{fig:model}
\end{figure}

Contribution (a) in Fig.~\ref{fig:model} is
the photon correction to the $\pi^+\pi^-$ loop, with the pion mass fixed at its
isospin-symmetric value. We adopt the Edinburgh consensus (EC) scheme~\cite{FLAG:2024oxs}: in the
isospin-symmetric theory ($\alem=0$, $m_u=m_d$) all pions have $m_\pi^{\rm iso}=135.0$~MeV, and QED,
switched on with $m_u=m_d$ and the hadronic scale (here the $\rho$, $\omega$ and $f_\pi$ parameters)
held fixed, raises $m_{\pi^\pm}$ to its physical value but leaves the neutral pion essentially
unaffected. The conversion term (b) accounts for this shift of the charged-pion mass and hence of
the two-pion threshold. Assuming that the leading pion-mass
dependence comes from the two-pion states, we evaluate it as
\begin{equation}
\Delta\overline\Pi^{({\rm ct})}(Q^2)\simeq\overline\Pi^{(\pi\pi)}(Q^2)\big|_{m_\pi=m_{\pi^\pm}}
-\overline\Pi^{(\pi\pi)}(Q^2)\big|_{m_\pi=m_\pi^{\rm iso}},
\label{eq:ct}
\end{equation}
where $\overline\Pi^{(\pi\pi)}$ denotes the isospin-symmetric two-pion component of the
scalar hadronic vacuum polarization function $\overline\Pi$, which includes the electromagnetic coupling and
is renormalized to vanish at zero momentum, and $Q^2$ is the Euclidean momentum squared.
Contribution (c) comes from the $\pi^0\gamma$ intermediate state and is governed by the anomalous pion transition form factor.

\begin{figure}[!t]
\centering
\includegraphics[width=0.54\textwidth,clip]{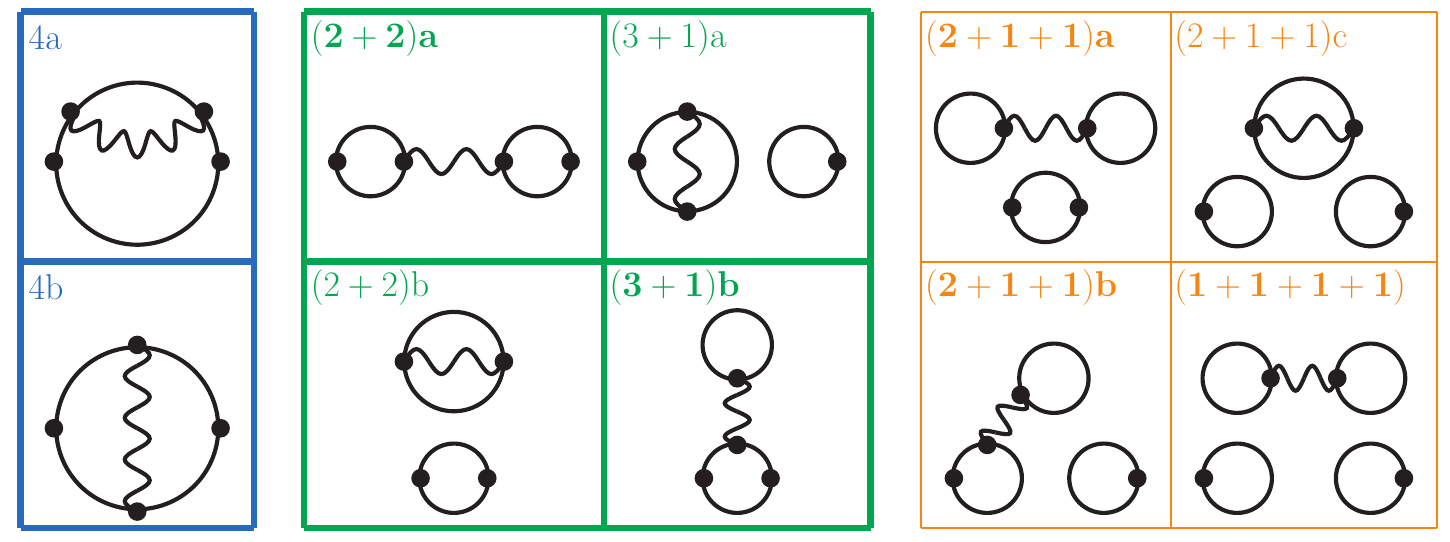}
\caption{Wick contractions of the $O(\alem)$ correction to the HVP. Solid lines: quark
propagators; wavy lines: photon; dots: insertions of the electromagnetic current. The
contractions labeled in bold are UV finite~\cite{Parrino:2025afq}.}
\label{fig:topologies}

\begingroup
\captionof{table}{LO PQChPT coefficients of Eq.~\eqref{eq:matching} for the light-quark contributions. Each
row sums to one; the $(2+1+1)$ and $(1+1+1+1)$ contractions vanish at this order.}
\label{tab:coeff}
\small
\renewcommand{\arraystretch}{1.15}
\begin{tabular}{lccccc}
\toprule
$(i)$ & 4a+4b & (2+2)a & (2+2)b & (3+1)a & (3+1)b \\
\midrule
$C^{(\pi\pi\gamma)}_{(i)}$ & $34/81$ & $50/81$ & $25/81$ & $-14/81$ & $-14/81$ \\
$C^{(\pi\gamma)}_{(i)}$ & $34/9$ & $-25/9$ & 0 & 0 & 0 \\
\bottomrule
\end{tabular}
\par
\endgroup
\medskip

\begingroup
\captionof{table}{Model predictions ($10^{-10}$; infinite volume, EC scheme) for (a)--(c)
and, via Eq.~\eqref{eq:matching}, the Wick contractions (UV-finite in bold); the negligible SD window ($<0.02$) is not shown.}
\label{tab:pred}
\small
\begin{tabular}{lrrrrrr}
\toprule
window [fm] & 0.4--1.0 & 1.0--1.5 & 1.5--2.0 & 2.0--2.8 & 2.8--$\infty$ & total\\
\midrule
(a) $\pi^+\pi^-\gamma$ & $-$0.1 & $+$0.1 & $+$0.2 & $+$0.3 & $+$0.3 & $+$0.7 \\
(b) conversion & $-$0.4 & $-$1.2 & $-$1.6 & $-$2.3 & $-$2.5 & $-$7.9 \\
(c) $\pi^0\gamma$ & 0.0 & 0.0 & 0.0 & 0.0 & 0.0 & $+$0.1 \\
\midrule
4a+4b & $-$0.2 & $-$0.4 & $-$0.5 & $-$0.7 & $-$0.8 & $-$2.6 \\
\textbf{(2+2)a} & $-$0.3 & $-$0.7 & $-$1.0 & $-$1.3 & $-$1.4 & $-$4.7 \\
(2+2)b & $-$0.1 & $-$0.3 & $-$0.4 & $-$0.6 & $-$0.7 & $-$2.2 \\
(3+1)a & $+$0.1 & $+$0.2 & $+$0.2 & $+$0.3 & $+$0.4 & $+$1.2 \\
\textbf{(3+1)b} & $+$0.1 & $+$0.2 & $+$0.2 & $+$0.3 & $+$0.4 & $+$1.2 \\
\midrule
total & $-$0.5 & $-$1.1 & $-$1.4 & $-$1.9 & $-$2.2 & $-$7.1 \\
\bottomrule
\end{tabular}
\par
\endgroup
\end{figure}

\section{Lattice matching and results}
\label{sec:lattice}

In the time-momentum representation~\cite{Bernecker:2011gh} the HVP contribution is the
Euclidean correlator of the current $J_\mu=\sum_f\mathcal{Q}_f\,\bar q_f\gamma_\mu q_f$ ($\mathcal{Q}_f$ in units
of $e$), folded with a known QED kernel~\cite{DellaMorte:2017dyu},
\begin{align}
\amu^{\rm HVP}&=\Big(\frac{\alem}{\pi}\Big)^2\int_0^\infty\!\dd t\,f(t)\,C(t),\qquad
C(t)=-\frac13\sum_{k=1}^3\int\!\dd^3x\,\big\langle J_k(t,\vec x)J_k(0)\big\rangle,
\label{eq:TMR}
\end{align}
where $f(t)\simeq\pi^2m_\mu^2t^4/9$ at small $t$ (this leading term cancels in the recently proposed mass-rescaled
muon-electron $(g-2)$ difference~\cite{Li:2026bgj}). The model enters through
the dispersion relation: the correlator follows from the imaginary part of the vacuum polarization,
\begin{equation}
C(t)=\frac{1}{24\pi^2}\int_{s_{\rm thr}}^\infty\!\dd s\,\sqrt s\,R(s)\,e^{-\sqrt s\,t},\qquad
R(s)=-\frac{3}{\alem}\,\im\overline\Pi(s).
\label{eq:Cdisp}
\end{equation}
Here $s_{\rm thr}$ is the lowest threshold. To separate distance scales we use the windows
\begin{equation}
\amu^{\rm win}=\Big(\frac{\alem}{\pi}\Big)^2\int_0^\infty\!\dd t\,f(t)\big[\Theta(t-t_1)-\Theta(t-t_2)\big]C(t),
\label{eq:windows}
\end{equation}
with $\Theta(t)=\frac12[1+\tanh(t/\Delta)]$ and $\Delta=0.15$~fm. The standard short-distance (SD),
intermediate (ID) and long-distance (LD) windows are separated at $t=0.4$ and
$1.0$~fm; for SD (LD), the lower (upper) term is replaced by 1 (0).

On the lattice the $O(\alem)$ correction to $C(t)$ splits into the Wick contractions of
Fig.~\ref{fig:topologies}. Those in which the photon joins two different quark loops are UV
finite~\cite{Parrino:2025afq}; the largest, the ``dumbbell'' $(2+2)$a, is the natural target of a
scheme-independent lattice determination. In the others, 4a, 4b, $(2+2)$b, $(3+1)$a and $(2+1+1)$c, the
photon is attached twice to the same quark loop, so their bare contributions are UV divergent and require counterterms.

For properly renormalized currents, counterterms cancel these UV
divergences~\cite{Biloshytskyi:2022ets}. For a lattice comparison, their finite parts are fixed
by matching pure QCD at the EC inputs to physical QCD+QED, with strong
isospin breaking treated separately. In this convention, 4a+4b includes the
connected quark-mass insertions, $(3+1)$a the valence-disconnected
ones, and $(2+2)$b the sea-quark-mass and strong-coupling insertions into the connected HVP.
The one-photon-reducible part of $(2+2)$a is assigned to the NLO HVP~\cite{Beltran:2026nlo}.

LO PQChPT charge factors for $\pi^0$ exchange and charged-pion loops
carry over from light-by-light scattering once the photon-connected currents are
specified, as shown for $(2+2)$a in Ref.~\cite{Parrino:2025afq}. The model estimate for quark contraction $(i)$ reads
\begin{equation}
\overline\Pi_{(i)}=C^{(\pi\pi\gamma)}_{(i)}\big[\overline\Pi^{(\pi\pi\gamma)}+\Delta\overline\Pi^{({\rm ct})}\big]
+C^{(\pi\gamma)}_{(i)}\,\overline\Pi^{(\pi^0\gamma)},
\label{eq:matching}
\end{equation}
where $\overline\Pi^{(\pi\pi\gamma)}$ and $\overline\Pi^{(\pi^0\gamma)}$ denote contributions (a) and (c), with the
coefficients of Table~\ref{tab:coeff}. The conversion term
shares the $\pi^+\pi^-\gamma$ coefficients, as the pion-mass shift has the flavor structure of a
photon within the charged-pion loop. This identification is well defined for UV-finite
contractions; the others rely on the convention above, which the model approximates by
fixing the isospin-symmetric pion mass in (a) and adding the conversion term (b).

Table~\ref{tab:pred} gives infinite-volume model predictions for individual Wick contractions.
We split the LD window at 1.5, 2.0 and 2.8~fm, typical cutoffs beyond which lattice collaborations reconstruct the
correlator by other methods.
Although all three contributions are formally $O(\alem)$, the conversion term (b) is one order of magnitude larger than (a) and (c) and dominates the model sum, $-7.1\tenten$.
The UV-finite dumbbell contributes $-4.7\tenten$, about two thirds of the model sum, compatible with
lattice results $(-6.9\pm2.9)\tenten$~\cite{RBC:2018dos} and
$(-6.4\pm1.3)\tenten$~\cite{Parrino:2025afq}. The SD contribution is negligible and the ID contribution small.
The window beyond 2.8~fm accounts for $\sim$30\% of each contraction in Table~\ref{tab:pred}.
Beyond 1.0~fm, the dumbbell/total ratio is $\sim 0.7$ in each window.

Additional $\pi\pi$ rescattering through the $\sigma$ resonance, roughly estimated in the linear $\sigma$
model with VMD pion form factors and $m_\sigma\simeq 500$~MeV, contributes about $-0.3\tenten$
to the full QED correction. Including the strange quark adds
$\eta\gamma$, $\eta'\gamma$ and kaon-loop contributions of $O(10^{-11})$ or below, given the
VMD parameters of Ref.~\cite{Parrino:2025afq}. Nevertheless, one should assign a very conservative 30\%
model uncertainty to the values in the ``total'' column of Tab.~\ref{tab:pred}, covering omitted inelastic channels, matching assumptions and chiral
behavior.

\section{Conclusions}
\label{sec:conclusions}

We provide phenomenological benchmarks for the light-quark Wick contractions and time
windows of the electromagnetic HVP correction using a VMD model matched to certain quark topologies via 
LO PQChPT. The pion-mass
conversion dominates the model sum, while about 30\% of the UV-finite dumbbell
lies beyond 2.8~fm. A comparison with BMW lattice data, including
finite-volume effects, is in preparation.

\begin{acknowledgement}
I thank L.~Lellouch, K.~Szabo, G.~Wang, D.~Erb, H.~B.~Meyer, J.~Parrino and V.~Pascalutsa for
collaboration on the topics presented here. This work was supported by the Excellence
Initiative of Aix-Marseille University---A$^{*}$Midex, a French ``Investissements d'Avenir'' program,
under grant~\mbox{AMX-22-RE-AB-052}, and by the French National Research Agency under
contract~\mbox{ANR-22-CE31-0011}.
Claude Opus 5.5 (Anthropic) and ChatGPT 6.1 Sol (OpenAI) were used for drafting and code editing;
everything was checked by the author.
\end{acknowledgement}

\begingroup
\interlinepenalty=10000 

\renewcommand{\texttt}[1]{\href{https://arxiv.org/abs/#1}{\nolinkurl{arXiv:#1}}}

\renewcommand{\doiwoc}[1]{{\urlstyle{same}\href{https://doi.org/#1}{\nolinkurl{#1}}}}
\bibliography{refs}
\endgroup

\end{document}